\documentclass[5p,numbers,sort&compress,preprint]{elsarticle}
\usepackage{graphicx,subfigure,amsbsy,amsmath,amssymb}
\begin{document}

\title{Temporal response of nonequilibrium correlated electrons}

\author[SIMES,UND]{B.~Moritz\corref{cor1}\fnref{email}}
\author[SIMES,Geballe]{T.~P.~Devereaux}
\author[Georgetown]{J.~K.~Freericks}

\cortext[cor1]{Corresponding author}  
\fntext[email]{E-mail address:  moritzb@slac.stanford.edu}

\address[SIMES]{Stanford Institute for Materials and Energy Science, SLAC National Accelerator Laboratory, Menlo Park, CA 94025, USA}
\address[UND]{Department of Physics and Astrophysics, University of North Dakota, Grand Forks, ND 58202, USA}
\address[Geballe]{Geballe Laboratory for Advanced Materials, Stanford University, Stanford, CA 94305, USA} 
\address[Georgetown]{Department of Physics, Georgetown University, Washington, DC 20057, USA}

\date{\today}

\begin{abstract}
In this work we examine the time-resolved, instantaneous current response for the spinless Falicov-Kimball model at half-filling, on both sides of 
the Mott-Hubbard metal-insulator transition, driven by a strong electric field pump pulse.  The results are obtained using an exact, nonequilibrium, 
many-body impurity solution specifically designed to treat the out-of-equilibrium evolution of electrons in time-dependent fields.  We provide a 
brief introduction to the method and its computational details.  We find that the current develops Bloch oscillations, similar to the case of DC 
driving fields, with an additional amplitude modulation, characterized by beats and induced by correlation effects.  Correlations primarily manifest 
themselves through an overall reduction in magnitude and shift in the onset time of the current response with increasing interaction strength.     
\end{abstract}


\maketitle

We examine the time-domain current response for the spinless Falicov-Kimball (FK) model at half-filling.  This model provides a relatively simple 
test-bed for understanding the physics of correlated electron systems.  In particular, the model possesses a Mott-Hubbard metal-insulator transition 
(MIT) at half-filling, as well as non-Fermi liquid behavior.~\cite{Freericks_FK}  The appeal of this model for the present study rests on the exact, 
nonequilibrium, many-body impurity solution that has been developed for the situation where the electrons are subject to time-dependent driving 
fields.~\cite{Freericks_1,Freericks_2,augsburg}  Here, the field, or pump pulse, that drives the system out of equilibrium is modeled as a large 
amplitude electric field with a characteristic Gaussian profile.  In principle the field may take any form, and we also choose to study fields 
modified through the inclusion of a harmonic modulation.  This additional modulation leads to noticeable changes in the time-resolved current 
response of the system.

Nonequilibrium dynamical mean-field theory (DMFT) is employed to obtain the real-time dynamics for the FK model on the hypercubic lattice in infinite 
dimensions ($d=\infty$).~\cite{Freericks_1,Freericks_2,augsburg}  The Hamiltonian in equilibrium takes the form
\[
H_{\rm eq} = -\frac{t^{*}}{2\sqrt{d}}\sum_{<ij>}(c^{\dag}_{i}c_{j} + h.c.) - \mu\sum_{i}c^{\dag}_{i}c_{i} + U\sum_{i}w_{i}\,c^{\dag}_{i}c_{i}.
\]
The FK model describes the hopping of conduction electrons that experience an on-site interaction $U$ with another species of localized electrons 
where $U_{c}=\sqrt{2}\,t^{*}$ is the critical interaction strength for the MIT at half-filling.  Throughout this work, the energy unit is taken to be 
$t^{*}$.  At half-filling, the average localized electron occupation $\left<w_{i}\right> = 0.5$ and the chemical potential $\mu = U/2$.  The 
nonequilibrium DMFT formalism proceeds in essentially the same manner as the iterative approach applied in equilibrium~\cite{Jarrell_DMFT} where all 
quantities now have two time indices.

The driving term is modeled by a spatially uniform electric field along the $(1, 1, 1,\ldots)$ hypercubic body diagonal.  This choice simplifies 
evaluation of the noninteracting Green's function (GF).~\cite{Freericks_1,Freericks_2}  The field takes the form 
\[
\mathbf{E}(t) = \mathbf{E}_{o}\cos(\omega(t-t_{o})+\phi)\exp^{-(t-t_{o})^{2}/\sigma^{2}}.
\]
The influence of this driving field enters through the associated vector potential $\mathbf{A}(t)$ in the Hamiltonian gauge using a Peierls' 
substitution~\cite{Peierls_sub}.  The time-resolved current response of the system is the velocity weighted average of the \emph{equal-time} lesser 
GF~\cite{Freericks_1,Freericks_2}
\[
\left<j_{i}(t)\right>  = -ie\sum_{\mathbf{k}}v_{i}(\mathbf{k}-e\mathbf{A}(t))G^{<}_{k}(t,t), 
\]  
where all velocity components $v_{i}$ are equal for the chosen driving field direction.  In the nonequilibrium DMFT formalism the sum over momentum 
can be converted to a two dimensional integral over two ``band energies" distributed with a Gaussian joint density of states on the hypercubic 
lattice.  The velocity components also can be expressed in terms of these band energies.~\cite{Freericks_2}  The continuous matrix operators of the 
nonequilibrium DMFT formalism are approximated by discretizing the Keldysh contour.  Quadratic extrapolation of results to zero ``step size" on the 
contour ensures that sum rules for the spectral moments are satisfied within a few percent for strong correlations and high fields.~\cite{sumrules}  
This quadratic extrapolation based on Lagrange interpolation also produces accurate results for current.

\begin{figure}[th!]
\centering\includegraphics[width=3.375in]{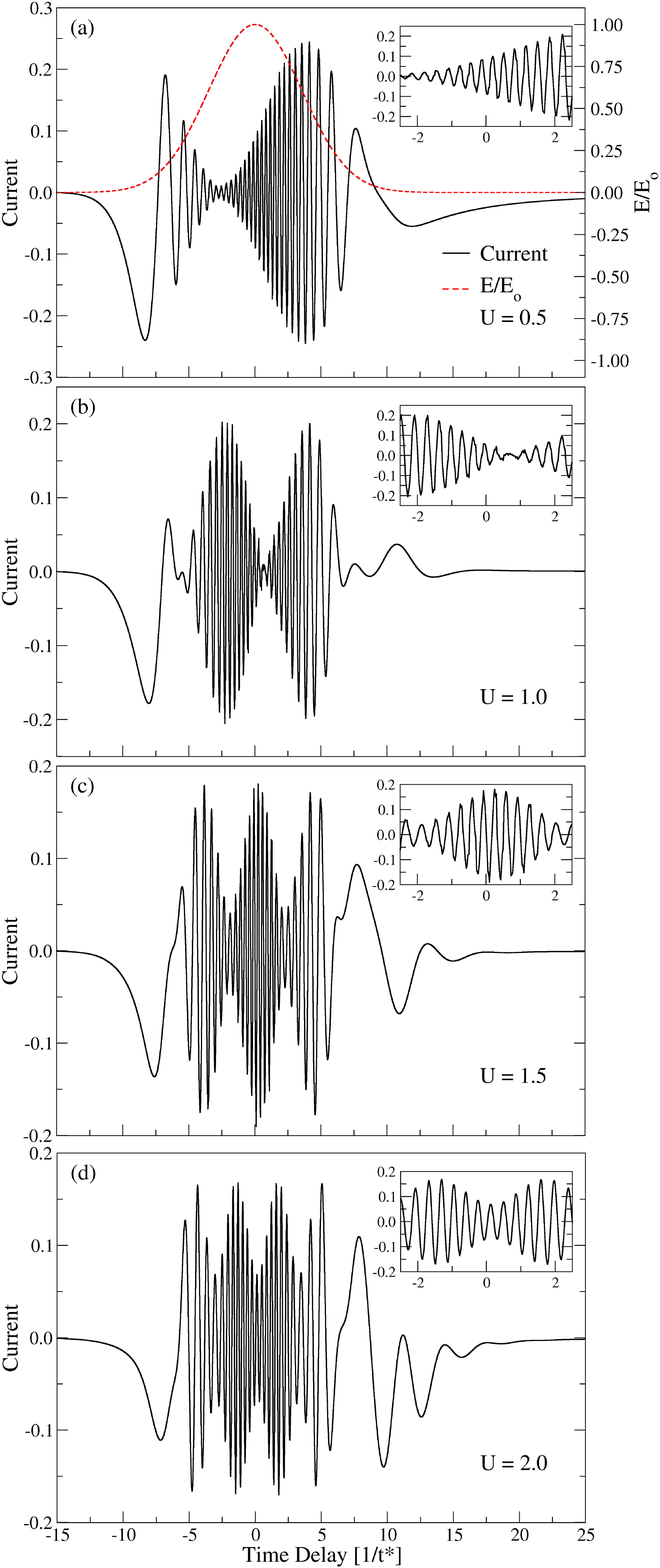} \caption{Time-resolved current response of the FK model with interaction strengths (a) $U 
= 0.5$, (b) $U = 1.0$, (c) $U = 1.5$, and (d) $U = 2.0$ for a system subject to an electric field pulse whose profile, plotted in arbitrary units, is 
shown in panel (a) ($E_{o} = 20$, $t_{o} = 0$, $\sigma = 5$, $\omega = 0$, $\phi = 0$).  The inset in each panel highlights the current response near 
the maximum in the field profile at a time delay $t = 0$ (between time delays $t = -2.5$ and $2.5$).  (Color online.)\label{fig:1}}
\end{figure}

\begin{figure}[th!]
\centering\includegraphics[width=3.375in]{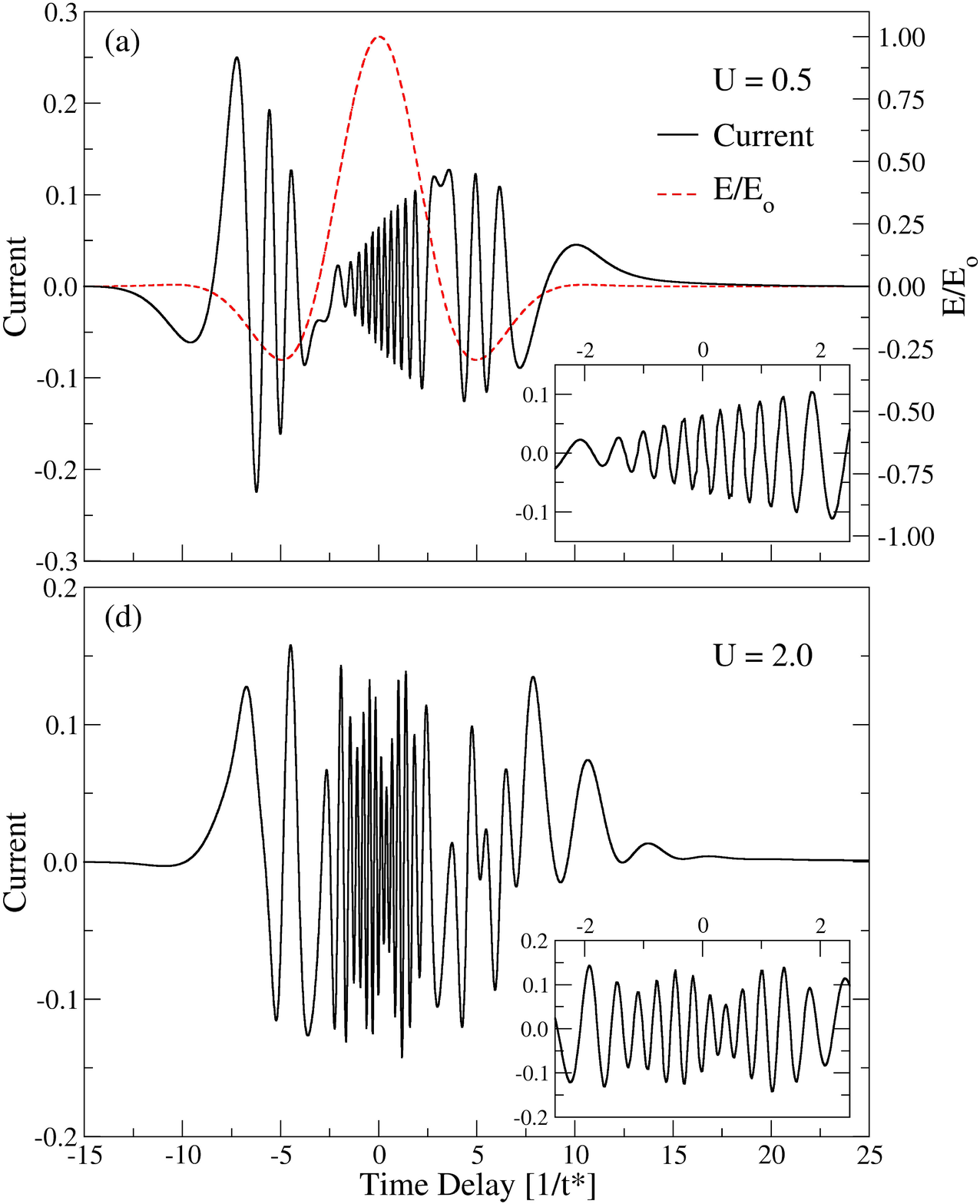} \caption{Time-resolved current response of the FK model with interaction strengths (a) $U 
= 0.5$ and (b) $U = 2.0$.  The profile of the electric field pulse, plotted in arbitrary units, is shown in panel (a) ($E_{o} = 20$, $t_{o} = 0$, 
$\sigma = 5$, $\omega = 0.5$, $\phi = 0$).  The inset in the two panels highlights the current response between time delays $t = -2.5$ and $2.5$.  
(Color online.)\label{fig:2}}
\end{figure}

The Gaussian distribution in both band energies makes Gaussian quadrature an ideal method for dealing with the two dimensional Hilbert transform that 
yields the local GF.  In this study the results for the local GF and local self-energy were obtained by averaging the Hilbert transform over 
quadratures with $N = 54$ and $N = 55$ points in each dimension for a total of 5,941 quadrature points.  This represents the most computationally 
intensive portion of the algorithm, but also provides an opportunity for parallelization as the evaluation of the integrand of the transform is 
independent for every quadrature point.  

In practice, the integrand for each quadrature point, obtained from a matrix inversion and two matrix multiplies, is a dense, complex matrix whose 
size depends on the discretization of the Keldysh contour.  For the cases considered here, reasonable accuracy for the spectral moments following 
quadratic extrapolation could be achieved using matrices ranging in size up to $4100\times4100$ elements.~\cite{Freericks_2}  This provides a rather 
severe ``bottleneck" to efficient scaling, using the master-slave model, if each slave were to communicate its results directly with the master.  
Instead, a recursive binary gather~\cite{Freericks_RBG} has been employed which involves a progressive binary division of the slaves and inter-slave 
communication to accumulate the results of the quadrature integration for the Hilbert transform.  Additional details about the numerical issues 
associated with this method may be found in Refs.~\cite{Freericks_RBG,Freericks_HPCMP_2005,Freericks_HPCMP_2007}.  Note that in this form the 
parallel code scales reasonably well on a number of different platforms~\cite{Freericks_2,Freericks_RBG,Freericks_HPCMP_2005,Freericks_HPCMP_2007} 
with the limitation coming from the serial portion of the code where the impurity solver determines the local, double-time self-energy.  The results 
presented from this study have been obtained from calculations performed on typically $\sim 1000 - 1500$ processors with $\sim 30 - 40$ 
nonequilibrium DMFT iterations to achieve convergence of the self-energy.

Figure~\ref{fig:1} shows the time-resolved current response for the FK model at half-filling, and for various interaction strengths, subjected to a 
pump pulse with parameters $E_{o} = 20$, $t_{o} = 0$, $\sigma = 5$, $\omega = 0$, and $\phi=0$.  Bloch oscillations are visible clearly in each 
panel, especially within the insets showing the current response near the center of the pump pulse at $t = 0$.  The frequency of these Bloch 
oscillations, proportional to the electric field strength $E$ for DC driving fields,~\cite{Freericks_1,Freericks_2} varies as a function of time due 
to changes in the field strength associated with the Gaussian profile of the pump pulse.  An additional amplitude modulation, characterized by beats 
in the current response, also appears with a frequency proportional to the interaction strength $U$.  The Bloch oscillations and amplitude modulation 
appear more regular for interactions strengths $U>U_{c}$ than for the case of a DC driving field.  This behavior likely results from the large 
amplitude of the Gaussian pump pulse that reduces the effective damping from electron correlations, especially near the center of the pump pulse.  
However, damping effects are manifest in the general reduction in the magnitude of the current response and a shift in the onset of the current 
response to higher time delays with increasing correlation strength, progressing from Fig.~\ref{fig:1}(a) to Fig.~\ref{fig:1}(d).

The results in Fig.~\ref{fig:2} show the response for a pump pulse with parameters $E_{o} = 20$, $t_{o} = 0$, $\sigma = 5$, $\omega = 0.5$, and $\phi 
= 0$.  The pump pulses in this study are assumed to be phase-locked (the phase $\phi$ does not vary from pulse to pulse or this could be viewed as 
the ``single-shot" response).  While one clearly sees Bloch oscillations, they appear far more irregular than those presented in Fig.~\ref{fig:1}.  
This is presumably due to the additional harmonic modulation of the pump pulse, especially for $U>U_{c}$ (see Fig.~\ref{fig:2}(b)).  For weak 
correlations, the onset of the current occurs in a region where field is strong enough, although still quite small compared to the maximum at $t = 
0$.  The additional amplitude modulation appears for $U<U_{c}$ in Fig.~\ref{fig:2}(a), but this behavior is apparent only within the relatively 
narrow central peak of the pump pulse.  The damping leads to a more irregular evolution of the Bloch oscillations with increasing correlation 
strength compared to the results in Fig.~\ref{fig:1}; however, there remains a general reduction in the magnitude and onset of the current response 
with increased correlations, even more pronounced than that shown in Fig.~\ref{fig:1}.

The results presented from the current model capture the formation of Bloch oscillations in the time-resolved current response of a correlated system 
subject to a particularly strong pump pulse with a Gaussian profile.  Of course, this behavior is quite similar to the results presented for DC 
driving fields,~\cite{Freericks_1,Freericks_2} but the field strength is greatly increased to ease issues associated with convergence and 
extrapolation in the present study.  The time-resolved current is one of the most readily accessible response functions that can be used to 
characterize the out-of-equilibrium behavior of a correlated electron system subject to time-dependent fields.  Using the same data generated from 
these calculations, one also can access the time-resolved photoemission response~\cite{Freericks_3,Moritz_tr-PES} that provides a more direct measure 
of the redistribution of spectral weight and rearrangement of electronic states that should characterize the nonequilibrium correlated system in a 
pump-probe experiment.

{\it Acknowledgements}. The authors would like to thank H. Craig, P. S. Kirchmann, H. R. Krishnamurthy, F. Schmitt, Z.-X. Shen, and M. Wolf for 
valuable discussions.  B.M. and T.P.D. were supported by the U.S. Department of Energy, Office of Basic Energy Sciences, under contract 
DE-AC02-76SF00515.  J.K.F. was supported by the U.S. Department of Energy, Office of Basic Energy Sciences, under grant number DE-FG02-08ER46542.  
Additional support was provided by the Computational Materials Science Network program of U.S. Department of Energy, Office of Basic Energy Sciences, 
Division of Materials Science and Engineering, under grant number DE-FG02-08ER46540.  The bulk of the computational results were made possible by an 
Innovative and Novel Computational Impact on Theory and Experiment (INCITE) award utilizing the resources of the National Energy Research Scientific 
Computing Center (NERSC) supported by the U.S. Department of Energy, Office of Science, under Contract No. DE-AC02-05CH11231.

\bibliographystyle{elsarticle-num}
\bibliography{Moritz_CCP_2009}

\begin{thebibliography}{10}
\expandafter\ifx\csname url\endcsname\relax
  \def\url#1{\texttt{#1}}\fi
\expandafter\ifx\csname urlprefix\endcsname\relax\def\urlprefix{URL }\fi
\expandafter\ifx\csname href\endcsname\relax
  \def\href#1#2{#2} \def\path#1{#1}\fi

\bibitem{Freericks_FK}
J.~K. Freericks, V.~Zlati\'{c}, Rev. Mod. Phys. 75 (2003) 1333--1382.

\bibitem{Freericks_1}
J.~K. Freericks, V.~M. Turkowski, V.~Zlati\'{c}, Phys.\ Rev.\ Lett. 97 (2006)
  266408.

\bibitem{Freericks_2}
J.~K. Freericks, Phys.\ Rev.\ B 77 (2008) 075109.

\bibitem{augsburg}
M.~Eckstein, M.~Kollar, Phys.\ Rev.\ B 78 (2008) 245113.

\bibitem{Jarrell_DMFT}
M.~Jarrell, Phys.\ Rev.\ Lett. 69 (1992) 168--171.

\bibitem{Peierls_sub}
R.~E. Peierls, Z.\ Phys. 80 (1933) 763--791.

\bibitem{sumrules}
V.~M. Turkowski, J.~K. Freericks, Phys.\ Rev.\ B 77 (2008) 205102.

\bibitem{Freericks_RBG}
J.~K. Freericks, V.~M. Turkowski, V.~Zlati\'{c}, in: D.~E. Post (Ed.),
  Proceedings of the HPCMP Users Group Conference 2006 (Denver, CO), IEEE
  Computer Society, Los Alamitos, CA, 2006.

\bibitem{Freericks_HPCMP_2005}
J.~K. Freericks, V.~M. Turkowski, V.~Zlati\'{c}, in: D.~E. Post (Ed.),
  Proceedings of the HPCMP Users Group Conference 2005 (Nashville, TN), IEEE
  Computer Society, Los Alamitos, CA, 2005.

\bibitem{Freericks_HPCMP_2007}
J.~K. Freericks, in: D.~E. Post (Ed.), Proceedings of the HPCMP Users Group
  Conference 2007 (Pittsburgh, PA), IEEE Computer Society, Los Alamitos, CA,
  2007.

\bibitem{Freericks_3}
J.~K. Freericks, H.~R. Krishnamurthy, T.~Pruschke, Phys.\ Rev.\ Lett. 102
  (2009) 136401.

\bibitem{Moritz_tr-PES}
B.~Moritz, T.~P. Devereaux, J.~K. Freericks, arXiv:0908.1807v1
  [cond-mat.str-el].

\end{thebibliography}

\end{document}